# Quantum-Inspired and Optimization-Based CT Reconstruction Through Spin-Error Mitigation

Hyunju Lee[1,2], Jeonghwa Lee[2], Yushuang Sam Yang[3,*], Clement Chu[3], Anthony B. Murphy[4], and Kyungtaek Jun[2,5,*]

[1] Department of Mathematics, Dongguk University, Seoul, 04620, South Korea
[2] Quantum Research Center, QTomo, Chungcheongbuk-do, 28535, South Korea
[3]CSIRO, Research Way, Clayton, Victoria 3168, Australia
[4]CSIRO, PO Box 218, Lindfield, NSW 2070, Australia
[5]Chungbuk Quantum Research Center, Chungbuk National University, Chungcheongbuk-do, 28644, South Korea

***Corresponding author:** Kyungtaek Jun (ktfriends@gmail.com) and Yushuang Sam Yang (sam.yang@csiro.au).

**Abstract**

Quantum-optimization-based computed tomography (CT) reconstruction has demonstrated quantum supremacy in reconstruction accuracy over conventional classical methods under sparse-view and corrupted-sinogram conditions. However, scaling this approach remains challenging because larger images require increasingly many binary variables and interactions in the underlying quadratic unconstrained binary optimization (QUBO) model. Even with iterative region-of-interest (ROI) refinement, a 60 × 60 ROI with three X-ray mass attenuation coefficients requires 10,800 binary variables under our encoding. As an initial step toward larger quantum CT images, we investigate simulated annealing (SA) followed by spin-error mitigation for optimisation (SEMO), without altering the CT-QUBO objective. We evaluate three 54-view ROI reconstruction instances and nine-view reconstruction with detector-row corruption, comparing SA–SEMO with selected hybrid quantum-classical solver results. For the ROI cases, SEMO reduced the numbers of incorrectly reconstructed pixels in low-budget SA outputs from 13 to 1, 38 to 0, and 54 to 15. With 20% detector-row corruption, three of four selected SA–SEMO pairs achieved exact reconstruction. Applying SEMO to one hybrid-solver output reduced its ROI error from 78 to 1 pixel. Improvements were instance-dependent: in one corrupted-sinogram case, QUBO energy decreased while image-domain error increased slightly; in one ROI case, the refined image retained 15 incorrect pixels despite having lower QUBO energy than ground truth. At the evaluated problem sizes, SA–SEMO attained reconstruction accuracy comparable to or better than selected unmitigated hybrid-solver results, providing an initial solution-quality benchmark for future larger-scale, parallelized quantum CT image reconstruction.

## 1. Introduction

Quantum annealing provides a framework for addressing discrete optimization problems formulated as quadratic unconstrained binary optimization (QUBO) models [1]. Quantum-optimization-based computed tomography (CT) reconstruction applies this framework to represent discrete image intensities using binary variables and reconstruct images by minimizing a projection-consistency objective [2]–[4]. In selected sparse-view and artifact-corrupted benchmarks, quantum CT image reconstruction algorithms have demonstrated quantum supremacy in reconstruction accuracy over conventional classical reconstruction methods [2]–[4]. From a mathematical perspective, the projection-consistency objective corresponds to a binary least-squares formulation of the discretized CT linear system, providing a direct connection to QUBO formulations for systems of linear equations [5]. In this study, we refer to images reconstructed using our CT-QUBO formulation as quantum CT images, regardless of whether the underlying optimization is performed using a quantum annealer, a hybrid quantum-classical solver, or classical simulated annealing (SA). Despite these promising reconstruction results, increasing the spatial extent and resolution of these images remains challenging. Quantum processing units (QPUs) provide finite hardware resources and restricted qubit connectivity, while dense interactions in CT-derived QUBOs can require substantial additional resources for minor embedding [6]. Although hybrid solvers accommodate larger optimization models, increasing the numbers of binary variables and pairwise interactions still imposes considerable computational demands [7], [8].

Previous studies have addressed reconstruction-scale limitations through different strategies. Iterative reconstruction is well established in conventional CT [9], while recent quantum-assisted approaches have introduced coarse-to-fine reconstruction and region-wise refinement to reduce the number of binary variables optimized simultaneously [10]. Nevertheless, the reconstruction domain remains restricted in practice. For example, under the adjacent-level difference encoding used in our CT-QUBO formulation, a $60 \times 60$ region of interest (ROI) with three X-ray mass attenuation coefficients, $\{0, 1, 2, 3\}$, uses 10,800 binary variables (three per pixel). This is the largest ROI size evaluated in the present experiments, including the corresponding hybrid-solver results. Further enlargement substantially increases the optimization workload because projection-domain data fidelity introduces extensive interactions among image variables. In addition, physical quantum annealers have relatively few qubits, and coefficient dynamic-range and hardware-precision limitations can affect sampled-solution quality [11]. Consequently, extending quantum CT image reconstruction requires not only reducing the optimization domain but also investigating alternative computational strategies.

One possible direction is to employ software-based optimization, where the logical QUBO can be processed without directly embedding the full problem onto a physical QPU. SA provides such a quantum-inspired approach and offers opportunities for parallel implementation and numerical control [12]. However, SA may still return imperfect low-energy configurations, particularly under limited computational budgets. Spin-error mitigation for optimisation (SEMO), recently investigated by Yang et al. in quantum-annealing-based three-dimensional image segmentation [13], [14] provides a post-sampling approach for refining candidate solutions. In contrast to hardware-level qubit error correction QEC [15], quantum error mitigation (QEM) aims at reducing the impact of errors in quantum computing without the active feedback correction of QEC. QEM offers an alternative set of techniques to mitigate, rather than actively protect against or remove at scale, the effects of these errors on quantum information processing [16]. As SEMO operates on the logical optimization problem without modifying the original QUBO or the upstream optimizer, it is applicable to quantum-inspired computations such as SA. When applied to SA-generated configurations, error mitigation refers to post-sampling refinement of candidate solutions rather than correction of physical qubit errors. Combining SA with SEMO may therefore offer a route to improving reconstruction quality while retaining the flexibility required for future software-based scaling.

Unlike the previously studied image-segmentation problem, CT reconstruction involves projection-domain interactions and may rely on sparse or corrupted measurements. Consequently, reducing QUBO energy does not necessarily guarantee improved agreement with the ground-truth image. Both optimization energy and image-domain reconstruction accuracy must therefore be considered when evaluating post-sampling refinement.

Motivated by these considerations, this study investigates SA combined with SEMO as an initial step toward larger-scale quantum CT image reconstruction. We first evaluate three 54-view ROI-QCSTR problems, each containing 10,800 binary variables, and examine whether SEMO can improve low-budget SA solutions to achieve reconstruction accuracy comparable to existing hybrid-solver benchmarks. We then extend the investigation to sparse-view reconstruction with sinogram corruption that can induce ring artifacts. SEMO is additionally applied to selected hybrid-solver outputs. Particular attention is given to both reconstruction improvements and cases in which a lower QUBO energy does not correspond to higher image-domain accuracy.

The longer-term objective is to extend this software-based workflow through parallel computation, efficient QUBO representations, and compiled C/CUDA implementations, ultimately investigating reconstruction problems involving 100,000 or more binary variables. The present work serves as an initial validation of the SA–SEMO approach rather than a demonstration of such large-scale capability.

The remainder of this paper is organized as follows. Section 2 describes the CT-QUBO formulations, experimental settings, and SEMO implementation. Section 3 presents the reconstruction results and additional objective-function analysis. Section 4 discusses the findings and future scalability considerations, and Section 5 concludes the paper.

## 2. Experimental Methods

### 2.1 Study design and workflow

This study investigates SA combined with SEMO as an initial step toward a software-based framework for larger-scale quantum CT image reconstruction. Rather than introducing a new CT-QUBO formulation, we evaluate whether post-sampling SEMO can improve low-budget SA outputs to achieve reconstruction accuracy comparable to that obtained using higher-budget SA or the D-Wave hybrid solver. The present experiments focus on establishing the solution-quality performance of this workflow for existing CT-QUBO benchmarks, including 10,800-variable ROI instances, before investigating scalability to substantially larger reconstruction problems.

In CT reconstruction, projection data acquired from multiple viewing angles are assembled into a sinogram. In a CT-QUBO formulation, a candidate image encoded by binary variables is forward-projected to generate a predicted sinogram. The QUBO objective is then constructed by penalizing the discrepancy between the measured sinogram and the predicted sinogram generated from the candidate image.

Two CT-QUBO problem classes are used as experimental testbeds: classical compressed sensing (CCS)-based region-of-interest (ROI) quantum compressed sensing tomographic reconstruction (QCSTR) and ring-artifact QUBO reconstruction. These testbeds were previously constructed using projection-domain QUBO formulations, and they are reused here to evaluate the post-sampling behavior of SA and SEMO under different reconstruction conditions.

For each QUBO instance, SA was first applied under multiple sampling budgets and random seeds. The primary ROI comparison used a fixed low-budget SA configuration (20 reads, 50 sweeps, and seed 42), whereas the corrupted-sinogram experiments evaluated selected SA outputs under the settings specified in Table IV. These experiments were designed as paired, instance-level evaluations of SEMO rather than estimates of its overall success probability across random initializations. Selected low-budget SA solutions were then used as inputs to SEMO. The SEMO-mitigated solutions were evaluated against the ground-truth image and compared with solutions obtained using higher-budget SA and the D-Wave hybrid quantum-classical solver. This experimental design allows us to distinguish the quality of the initial SA sample from the additional improvement achieved through SEMO.

### 2.2 CT-QUBO formulation with sinogram consistency and spatial regularization

All CT-QUBO instances used in this study are based on a sinogram-consistency objective [2]. Let $P(\theta, s)$ denote the measured sinogram, where $\theta$ is the projection angle and $s$ is the detector position. For a candidate image whose pixel intensities are represented by binary decision variables $\vec{q}$, let $IP(\theta, s; \vec{q})$ denote the predicted sinogram obtained by forward-projecting the candidate image.

The sinogram-consistency term, which serves as the data-fidelity component of the QUBO, penalizes the difference between the measured sinogram and the predicted sinogram, with the constant term independent of the binary decision variables explicitly subtracted:

$$\sum_{\theta}\sum_{s}[P(\theta,s) - IP(\theta,s;\vec{q})]^2 - \sum_{\theta}\sum_{s}P(\theta,s)^2. \tag{1}$$

Following the least-squares-to-QUBO formulation for linear systems [5], the constant term is explicitly removed without affecting the minimizer. Since the predicted sinogram is linear in the binary variables and $q_i^2 = q_i$ for $q_i \in \{0, 1\}$, expanding the squared term in (1) yields a quadratic objective consisting of linear and pairwise interaction terms. Thus, the sinogram-consistency energy is equivalently written as

$$E_{sino}(\vec{q}) = \sum_{\theta}\sum_{s} IP(\theta,s;\vec{q})^2 - 2\sum_{\theta}\sum_{s}P(\theta,s)IP(\theta,s;\vec{q}). \tag{2}$$

Thus, minimizing the sinogram-consistency term favors a discrete candidate image whose predicted sinogram agrees with the measured sinogram used in the corresponding experiment.

Following the adjacent-level difference encoding of mass attenuation coefficients introduced in [17], each pixel is represented by three binary decision variables. For the intensity levels $\alpha_0 = 0, \alpha_1 = 1, \alpha_2 = 2$, and $\alpha_3 = 3$ used in this study, all adjacent-level differences are equal to one. Thus, the intensity of pixel $i$ is expressed as

$$x_i = q_{i,1} + q_{i,2} + q_{i,3}, \quad q_{i,k} \in \{0, 1\}. \tag{3}$$

Therefore, $x_i \in \{0, 1, 2, 3\}$, with the background level 0 represented by the all-zero state. For a $60 \times 60$ ROI, this representation results in $60 \times 60 \times 3 = 10{,}800$ binary decision variables.

Following the QCSTR formulation in [3], a quadratic spatial regularization term based on squared intensity differences between neighboring pixels is additionally incorporated. Let $\mathcal{N}$ denote the set of valid horizontal and vertical neighboring pixel pairs within the reconstruction region, with each pair counted once. The term is defined as

$$E_{CS}(\vec{q}) = \sum_{(i,j)\in\mathcal{N}} \left(x_i - x_j\right)^2, \tag{4}$$

where $x_i$ and $x_j$ denote the pixel intensities represented by the binary decision variables defined above. Since each pixel intensity is a linear function of the binary variables, $E_{CS}(\vec{q})$ is a quadratic function of $\vec{q}$.

Combining the sinogram-consistency and spatial regularization terms, the final QCSTR objective is expressed as

$$E(\vec{q}) = aE_{sino}(\vec{q}) + bE_{CS}(\vec{q}), \tag{5}$$

or equivalently, in QUBO form

$$E(\vec{q}) = \vec{q}^{\top} Q \vec{q}, \tag{6}$$

where $\vec{q}$ is the vector of binary decision variables and $Q = aQ_1 + bQ_2$ is the final QUBO matrix. Here, $Q_1$ and $Q_2$ denote the QUBO matrices corresponding to the sinogram-consistency and quadratic spatial regularization terms, respectively. The nonnegative weights $a$ and $b$ are fixed according to the CT-QUBO instance used in each experiment.

### 2.3 CT-QUBO test problems

Two classes of quantum CT image reconstruction problems were used as testbeds: CCS-based ROI-QCSTR and ring-artifact QUBO reconstruction. Both are formulated using the sinogram-consistency principle in Section 2.2, but they differ in the sinogram data used to construct the QUBO objective.

The ROI-QCSTR experiments used three discrete-valued 180 × 180 images: one synthetic phantom (ROI Sample 1) and two chest CT-derived phantoms (ROI Samples 2 and 3). The chest CT images originated from the TCGA-LUAD collection available through The Cancer Imaging Archive (TCIA) [18]. The source images for ROI Samples 2 and 3 were taken from the first DICOM files of the series listed at indices 83 and 75, respectively, in the locally generated series list. ROI Samples 2 and 3 were selected to provide structurally distinct regions with meaningful variation in discrete intensity levels. Each selected 512 × 512 CT slice was reduced to 170 × 170 by 3 × 3 block averaging and zero-padded by five pixels on each side to obtain a 180 × 180 image. Intensities were rescaled to a maximum of three and rounded to discrete levels, with background and non-target regions represented by zero. Thus, all three test images used a set of three X-ray mass attenuation coefficients, $\{0, 1, 2, 3\}$. A fixed 60 × 60 ROI was extracted from each image, yielding 3,600 ROI pixels and 10,800 binary variables using the encoding defined in Section 2.2.

#### 2.3.1 CCS-based ROI-QCSTR

The first test problem is based on a two-stage ROI-QCSTR framework using classical compressed sensing (CCS) for coarse reconstruction. In the first stage, CCS reconstructs a coarse global image from the sparse-view sinogram by imposing sparsity in the discrete cosine transform domain through an $l_1$-regularized reconstruction [19]. In the implementation used to construct the present test cases, the optimization was solved using the fast iterative shrinkage-thresholding algorithm (FISTA) [20]. The resulting coarse image provides a fixed estimate of the non-ROI background, while the second stage reconstructs only the selected ROI using the QCSTR QUBO. The sinograms were generated using the radon function in scikit-image, with 54 projection angles uniformly spaced from 0° to 180°, excluding 180°. The same angular configuration was used for the CCS coarse reconstruction and subsequent ROI-QUBO construction.

For ROI refinement, the non-ROI region is fixed using the coarse global image. Specifically, the ROI pixels of the coarse image are set to zero, producing the fixed non-ROI background image $I_{global}^{\backslash ROI}$. The sinogram generated from this fixed background is subtracted from the measured sparse-view sinogram $P(\theta, s)$. The residual sinogram is defined as

$$P_{res}(\theta, s) = P(\theta, s) - P_{I_{global}^{\backslash ROI}}(\theta, s), \tag{7}$$

where $P_{I_{global}^{\backslash ROI}}(\theta, s)$ denotes the sinogram generated from the fixed non-ROI background.

The residual sinogram $P_{res}(\theta, s)$ is then used as the measured sinogram term in the sinogram-consistency objective described in Section 2.2. The predicted sinogram term is generated only from the ROI image, whose pixel intensities are represented by binary decision variables. Thus, the ROI-specific sinogram-consistency energy, with the constant term explicitly subtracted, is expressed as

$$\sum_{\theta}\sum_{s}[P_{res}(\theta, s) - IP_{ROI}(\theta, s; \vec{q})]^2 - \sum_{\theta}\sum_{s} P_{res}(\theta, s)^2, \tag{8}$$

where $IP_{ROI}(\theta, s; \vec{q})$ denotes the predicted sinogram obtained by forward-projecting the binary-variable representation of the ROI. In this formulation, the non-ROI background is fixed, and only the ROI pixels are optimized.

The CCS-based ROI-QCSTR experiments were performed using three 54-view ROI cases. For each case, the ground-truth ROI was extracted from the corresponding quantum CT image with three X-ray mass attenuation coefficients, and the QCSTR weights were fixed at $a = b = 1$. The same QUBO instance was used to compare SA, SEMO-mitigated SA, and hybrid-solver solutions. The sample-specific ROI locations and QUBO weights are summarized in Table I.

Table I. CCS-based ROI-QCSTR test cases used in the 54-view experiments.

| Sample | ROI rows | ROI columns | ROI size | No. binary variables | QUBO weights |
|---|---|---|---|---|---|
| 1 | 67:127 | 60:120 | $60 \times 60$ | 10,800 | $a = 1, b = 1$ |
| 2 | 76:136 | 90:150 | $60 \times 60$ | 10,800 | $a = 1, b = 1$ |
| 3 | 70:130 | 60:120 | $60 \times 60$ | 10,800 | $a = 1, b = 1$ |

### 2.3.2 Ring-artifact QUBO reconstruction

The second test problem is ring-artifact QUBO reconstruction. In the earlier quantum CT image experiments [4], detector rows were randomly selected and perturbed using randomly generated corruption magnitudes to simulate detector defects. The resulting corrupted sinograms and corresponding artifact masks were saved and reused in the present study. For each corruption setting, the same saved data were used across the corresponding optimizer comparisons, keeping the corruption pattern fixed. The affected detector rows form stripe-like structures in the sinogram and can induce ring-shaped artifacts in the reconstructed image.

The ring-artifact experiments used a discretized Shepp–Logan phantom [21]. The 400 × 400 phantom provided by scikit-image was rescaled to 58 × 58, padded by one pixel on each side, and smoothed using a 7 × 7 Gaussian kernel. The resulting 60 × 60 image was normalized to a maximum intensity of three and rounded to obtain values in $\{0, 1, 2, 3\}$. Ten additional zero-valued pixels were padded on each side for forward projection.

To construct the QUBO, detector rows identified as corrupted were excluded from the sinogram-consistency objective rather than fitted as noisy measurements. Consequently, reconstruction was performed using the remaining unmasked projection data. Let $M(\theta, s)$ be a binary mask, where $M(\theta, s) = 1$ for clean sinogram entries and $M(\theta, s) = 0$ for corrupted or excluded entries. The masked sinogram-consistency energy, with the corresponding constant term explicitly subtracted, is given by

$$\sum_{\theta}\sum_{s} M(\theta, s)[P(\theta, s) - IP(\theta, s; \vec{q})]^2 - \sum_{\theta}\sum_{s} M(\theta, s)P(\theta, s)^2, \tag{9}$$

Here, $P(\theta, s)$ denotes the measured sinogram, and $IP(\theta, s; \vec{q})$ denotes the predicted sinogram generated from the candidate image represented by binary decision variables. Thus, only clean sinogram entries contribute to the QUBO construction, while corrupted detector rows are ignored during optimization. For both ring-artifact settings, the masked sinogram-consistency term in (9) was combined with the spatial regularization term in (4), using $a = b = 1$, to construct the final QCSTR objective in (5).

The main ring-artifact experiments considered two settings: 15% projection sampling with 20% detector-row corruption and 15% projection sampling with 30% detector-row corruption. Nine projection angles were uniformly spaced over [0°, 180°), corresponding to 15% of the angular sampling density of a 60-view reference. Both settings used the same phantom and projection angles, while the nominal fraction of corrupted detector rows was varied between 20% and 30%. The evaluated SA budgets and random seeds are reported separately for each setting in Table IV. The evaluation settings differed between the two corruption conditions. For 20% corruption, four selected SA–SEMO pairs with different sampling budgets and seeds were examined. For 30% corruption, the evaluation focused on two SA budgets, each repeated across four random seeds, after the initial results showed substantial residual reconstruction errors. These experiments were intended to assess the effect of SEMO within each fixed corruption setting rather than to isolate the effect of corruption level under matched optimization budgets.

### 2.4 Simulated annealing experiments

Simulated annealing (SA) was used to generate candidate solutions for the QUBO instances described in Section 2.3. For each instance, the QUBO matrix $Q$ was fixed, while the SA sampling budget and random seed were varied. This design produced annealing outputs with different levels of convergence, ranging from low-budget samples to stronger reference solutions.

For each sampled binary configuration $\vec{q}$, the QUBO energy was evaluated using the final QUBO matrix $Q$ defined in Section 2.2. Each SA output was decoded into a discrete image using the same three-variable sum encoding employed in the QUBO formulation, such that the three binary variables associated with each pixel were summed to obtain an intensity value in $\{0, 1, 2, 3\}$. The decoded image was then evaluated using the metrics described in Section 2.7.

Multiple SA budgets were evaluated using different combinations of reads and sweeps. For repeated-seed experiments, the random seeds were 42, 123, 124, and 125. SA-only results obtained with up to 300 reads and 1000 sweeps were retained as higher-budget references; these solutions were not assumed to be global QUBO optima.

For the primary 54-view ROI comparison, the initial SA setting was fixed at 20 reads, 50 sweeps, and seed 42 for each of the three instances. The corresponding saved SA outputs were used for the paired SEMO comparisons. Additional SA-only runs were analyzed separately to characterize reconstruction accuracy at larger sampling budgets.

### 2.5 Hybrid-solver implementation

The D-Wave Leap hybrid solver for binary quadratic models (BQMs) was used as an additional optimizer for selected CT-QUBO instances. For the 54-view ROI-QCSTR experiments, the previously constructed QUBO instances were submitted using `LeapHybridSampler` without specifying a custom solver identifier or an explicit `time_limit`, thereby using the service-provided default settings. Unlike direct quantum processing unit (QPU) sampling, this managed hybrid workflow does not require users to specify physical embedding, annealing schedules, or repeated QPU reads. The solver internally combines classical optimization with quantum-processing resources [7]. The lowest-energy configuration returned by the hybrid solver was selected for reconstruction evaluation and subsequent SEMO processing.

### 2.6 SEMO post-sampling error mitigation

SEMO was applied as a post-sampling error-mitigation method to candidate solutions obtained from SA and the hybrid solver. For each CT-QUBO instance, the same QUBO coefficients were used during the initial optimization and subsequent SEMO processing. SEMO was therefore applied without modifying the underlying CT reconstruction objective.

SEMO was applied to selected optimizer outputs using the corresponding QUBO coefficient dictionary. For the 54-view ROI-QCSTR experiments and the 30%-corruption ring-artifact experiments, the lowest-energy configuration from each selected SA run was supplied to SEMO as a single record, together with its QUBO energy and an occurrence count of one. For the 20%-corruption ring-artifact experiments, all records in each selected SA SampleSet were supplied, retaining their binary configurations, QUBO energies, and occurrence counts. For the hybrid-solver experiments, a selected lowest-energy configuration was supplied as a single record. The QUBO coefficients were loaded using SetupQUBO(Q), the error-mitigation method was configured using SetErrMitMethod(2, 1), and the data tolerance was set using SetDataTolerance(0). The SA solutions were passed to SEMO and processed using DoErrorMitigation with the SA-computed QUBO variable value configurations, energies, and occurrence counts.

The SEMO computation was performed on a Windows 11 Pro desktop computer equipped with an AMD Ryzen 9 9950X3D 16-Core Processor (4.30 GHz) and 96 GB of RAM. For each SA solution, a QUBO variable was selected randomly as the reference variable. An assembly of variable clusters was formed. A cluster can either be the reference variable, or the reference variable plus a variable coupled to the reference variable. Two variables were regarded as coupled when the QUBO quadratic coefficient between them was non-zero. Each such cluster was test-flipped to evaluate the impact on the objective function value. Flipping changes all variables in the cluster to different values; that is, when the original variable value in a cluster is 0, it is changed to 1 and vice versa. If any such test-flips reduced the objective function value, the flipped cluster that produced the maximum reduction was kept in the flipped state, and the objective function was updated accordingly. After completing computation for a reference variable, another reference variable was selected randomly, and the same process was repeated until each variable was selected twice as a reference variable in an SA solution. Each SA solution was processed using the above procedures until all SA solutions were processed.

After SEMO processing, the returned configurations were ranked by their QUBO energies, and the lowest-energy configuration was selected for reconstruction evaluation. The selected binary configuration was decoded into a discrete image using the three-variable sum encoding described in Section 2.2. The resulting image was compared with the corresponding ground-truth image using the metrics defined in Section 2.7.

The same post-sampling evaluation procedure was applied to selected outputs of the hybrid solver. Initial optimizer solutions and their SEMO-mitigated counterparts were evaluated using the same CT-QUBO objective and image-domain reconstruction metrics.

### 2.7 Reference solutions and evaluation metrics

The reconstruction quality of SEMO-mitigated solutions was evaluated against the corresponding ground-truth images with discrete intensity levels. In addition to the initial low-budget SA solutions, results obtained using higher-budget SA configurations and the hybrid solver were included as optimization references. These reference solutions were used to compare the reconstruction quality achieved by different optimization procedures; they were not assumed to represent globally optimal QUBO solutions.

For each reconstruction problem, the binary solution was decoded into a discrete image using the three-variable sum encoding described in Section 2.2. The decoded image was then compared pixelwise with the ground-truth image. Two image-domain metrics were considered: the number of incorrectly reconstructed pixels and mean absolute error (MAE).

Let $x_i^*$ and $\hat{x}_i$ denote the ground-truth and reconstructed intensity values at pixel $i$, respectively, and let $N$ denote the total number of pixels in the evaluated image or ROI.

The number of incorrectly reconstructed pixels was defined as

$$N_{\text{wrong}} = \sum_{i=1}^{N} \mathbf{1}\ (\hat{x}_i \neq x_i^*), \tag{10}$$

where $\mathbf{1}(\cdot)$ is the indicator function, which equals one when its argument is true and zero otherwise.

The MAE was calculated as

$$MAE = \frac{1}{N}\sum_{i=1}^{N} |\hat{x}_i - x_i^*|. \tag{11}$$

An exact reconstruction was defined as a reconstruction in which every reconstructed pixel intensity matched the corresponding value in the discrete ground-truth image, corresponding to $N_{\text{wrong}} = 0$.

The QUBO energy was also recorded to evaluate optimization-domain changes before and after SEMO processing. For each paired comparison, the lowest-energy initial optimizer solution and the lowest-energy SEMO-mitigated solution were evaluated using the same QUBO objective. The energy reduction was defined as

$$\Delta E = E_{\text{initial}} - E_{\text{SEMO}}, \tag{12}$$

where $E_{\text{initial}}$ and $E_{\text{SEMO}}$ denote the QUBO energies of the initial optimizer solution and the SEMO-mitigated solution, respectively. A positive value of $\Delta E$ therefore indicates a reduction in QUBO energy after SEMO processing.

Energy and image-domain metrics were evaluated separately because a reduction in QUBO energy does not necessarily imply a corresponding improvement in reconstruction accuracy, particularly when the objective is constructed using incomplete or corrupted projection data.

All image-domain metrics were calculated using the same ground-truth image and reconstruction region for each pair comparison. Consequently, changes in the reported metrics reflect differences between the initial optimizer output and the SEMO-mitigated solution for the same QUBO instance.

## 3. Results

### 3.1 SEMO error mitigation of low-budget ROI-QCSTR solutions

To evaluate SEMO under a controlled low-budget annealing condition, candidate solutions for the three ROI-QCSTR instances were generated using simulated annealing with the same setting of 20 reads, 50 sweeps, and random seed 42. This setting yielded solutions with residual reconstruction errors while keeping the initial SA budget and random seed fixed across the three reconstruction problems.

SEMO error mitigation reduced image-domain reconstruction errors in all three low-budget SA solutions. In Sample 1, the number of incorrectly reconstructed pixels decreased from 13 to 1, corresponding to a reduction in MAE from 0.003611 to 0.000278. In Sample 2, SEMO corrected all 38 incorrectly reconstructed pixels and achieved an exact reconstruction with zero MAE. In Sample 3, the number of incorrect pixels decreased from 54 to 15, while MAE decreased from 0.015000 to 0.004167. Thus, under an identical SA budget and random seed, SEMO reduced reconstruction error across all three ROI-QCSTR instances. Table II summarizes the paired quantitative comparisons, and Fig. 1 illustrates the corresponding reconstructed images.

**Table II. SEMO error mitigation for 54-view ROI-QCSTR reconstruction**

Each ROI contains 3,600 pixels. The initial SA solutions were obtained using 20 reads, 50 sweeps, and seed 42. Energy reduction is defined as $\boldsymbol{\Delta E = E_{SA} - E_{SEMO}}$ for each paired comparison; positive values indicate lower QUBO energy after SEMO processing.

| Sample | SA wrong pixels | SEMO wrong pixels | SA MAE | SEMO MAE | Energy reduction $\Delta E$ |
|---|---|---|---|---|---|
| 1 | 13 | 1 | 0.003611 | 0.000278 | 92.5 |
| 2 | 38 | 0 | 0.010556 | 0.000000 | 319.8 |
| 3 | 54 | 15 | 0.015000 | 0.004167 | 271.6 |

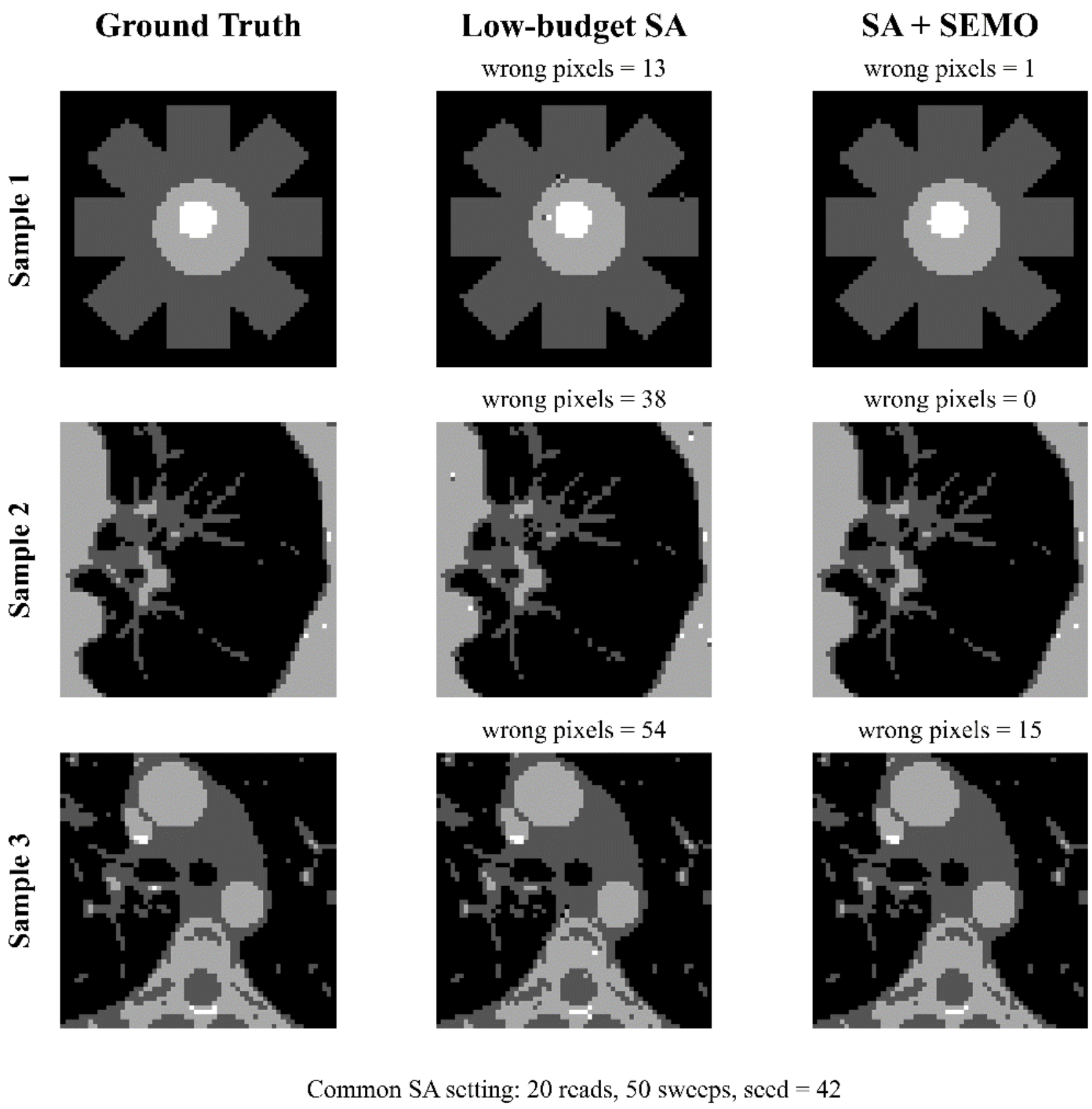

Figure 1. Comparison of ground-truth images, low-budget simulated annealing (SA) reconstructions, and SEMO-mitigated reconstructions for three 54-view ROI-QCSTR instances. The initial SA setting was fixed at 20 reads, 50 sweeps, and random seed 42 for all three instances. The number of incorrectly reconstructed pixels is indicated for each reconstructed image.

To investigate the 15 residual errors in ROI Sample 3, we compared the SEMO-mitigated configuration with the exact discrete ground truth under the same QCSTR objective. SEMO reduced the total QUBO energy by 271.560 relative to the initial SA solution. However, the SEMO reconstruction had a lower energy than the ground-truth configuration.

To quantify this difference, let $\boldsymbol{q}_{\mathrm{SEMO}}$ and $\boldsymbol{q}_{\mathrm{GT}}$ denote binary encodings of the SEMO-reconstructed and ground-truth ROIs, respectively. Using the sinogram-consistency and spatial-regularization terms defined in (2) and (4), their energy differences were calculated as

$$\Delta E_{\mathrm{sino}} = E_{\mathrm{sino}}(\vec{q}_{\mathrm{SEMO}}) - E_{\mathrm{sino}}(\vec{q}_{\mathrm{GT}}) = -21.067, \tag{13}$$

$$\Delta E_{\mathrm{CS}} = E_{\mathrm{CS}}(\vec{q}_{\mathrm{SEMO}}) - E_{\mathrm{CS}}(\vec{q}_{\mathrm{GT}}) = -3.000. \tag{14}$$

With $a = b = 1$, the total QCSTR energy of the SEMO reconstruction was therefore 24.067 lower than that of the ground-truth configuration. Thus, both objective terms assigned lower energies to the SEMO reconstruction than to the ground-truth configuration in this instance. All 15 residual pixel errors differed from the target by one intensity level. Restoring each incorrect pixel individually to its ground-truth intensity, while holding all other pixels fixed, increased the QUBO energy in every case (by 1.501–29.008). Restoring all 15 pixels simultaneously also increased the energy by 24.067.

### 3.2 Comparison with higher-budget simulated annealing

Table III compares the low-budget SA solutions before and after SEMO processing with SA-only solutions obtained using 300 reads and 1000 sweeps across four random seeds. For Samples 1 and 2, SEMO reduced the numbers of incorrectly reconstructed pixels from 13 to 1 and from 38 to 0, respectively, matching the higher-budget SA results for all four seeds. For Sample 3, SEMO reduced the number of incorrect pixels from 54 to 15, whereas higher-budget SA yielded 15, 13, 15, and 11 incorrect pixels for seeds 42, 123, 124, and 125, respectively. In a separate SA-only evaluation, Sample 1 also yielded one incorrect pixel using 100 reads and 100 sweeps for three of the four evaluated seeds.

Table III. Comparison of reconstruction accuracy between low-budget SA, SA followed by SEMO, and higher-budget SA for three 54-view ROI-QCSTR instances. Values indicate the number of incorrectly reconstructed pixels. Higher-budget SA results are reported in seed order: 42, 123, 124, and 125.

| Sample | Low-budget SA, (20 reads, 50 sweeps, seed 42) | Low-budget SA →SEMO | SA only, 300 reads, 1000 sweeps, seeds 42/123/124/125 |
|---|---|---|---|
| 1 | 13 | 1 | 1 / 1 / 1 / 1 |
| 2 | 38 | 0 | 0 / 0 / 0 / 0 |
| 3 | 54 | 15 | 15 / 13 / 15 / 11 |

### 3.3 SEMO error mitigation under sparse-view and corrupted-sinogram conditions

Under 15% projection sampling with 20% detector-row corruption, SEMO achieved exact reconstruction for a selected low-budget SA solution. Specifically, the number of incorrectly reconstructed pixels decreased from 38 to zero after SEMO error mitigation. Figure 2 illustrates the ground-truth phantom, the sparse and corrupted sinogram, and the reconstructed images before and after SEMO processing.

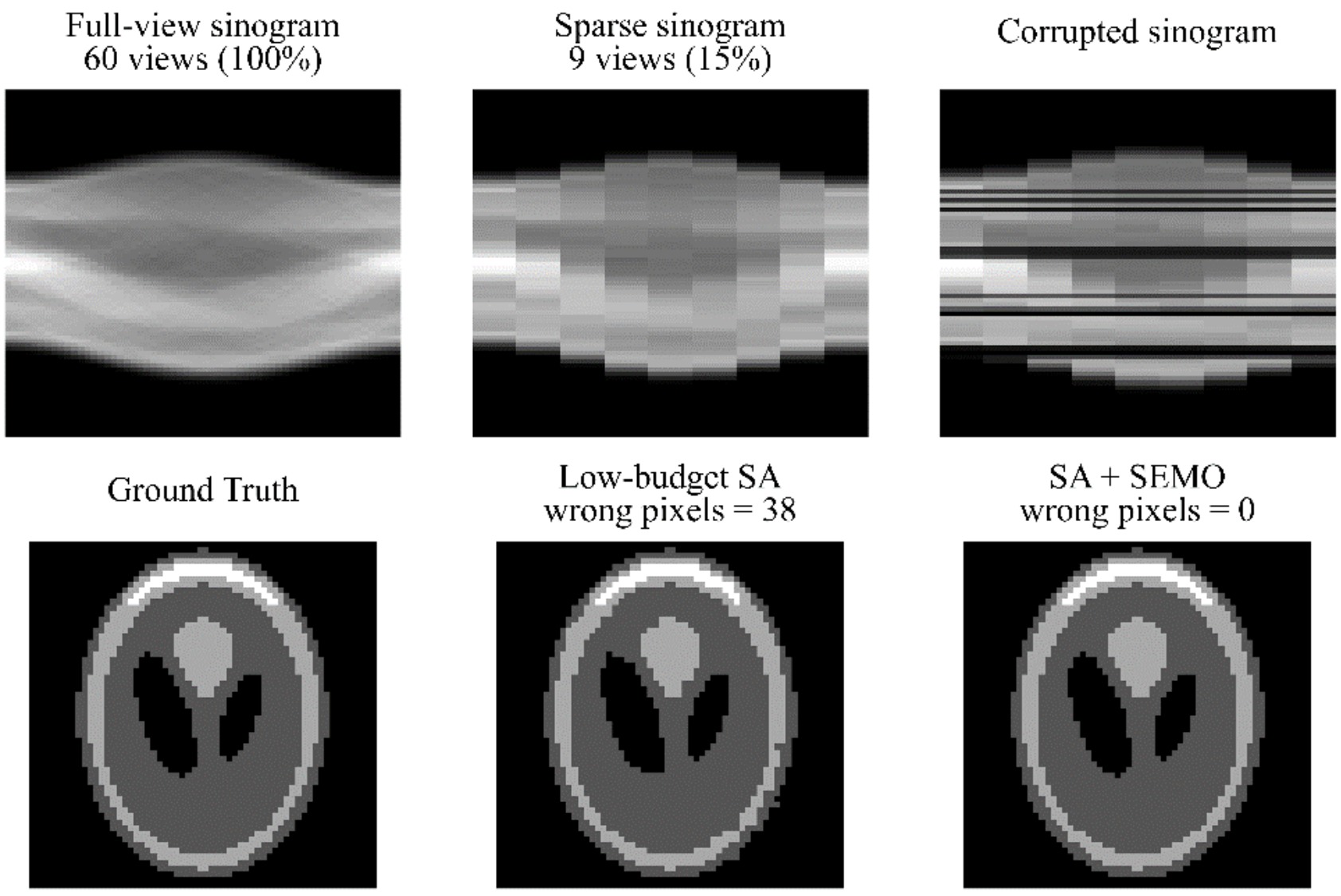

Figure 2. SEMO error mitigation in the ring-artifact reconstruction experiment with 15% projection sampling and 20% detector-row corruption. The upper row shows the full-view sinogram with 60 views, the sparse sinogram generated from nine uniformly spaced projection angles (15% of the 60-view reference angular count), and the corresponding corrupted sinogram. The lower row shows the ground-truth image, the low-budget SA reconstruction, and the SEMO-mitigated reconstruction. The number of incorrectly reconstructed pixels decreased from 38 to zero after SEMO processing.

Table IV summarizes the paired SA–SEMO results for the two detector-row corruption settings. The 20%-corruption results are reported individually for four selected runs with different combinations of SA budgets and random seeds. The 30%-corruption results are reported as mean ± sample standard deviation across four random seeds for each of two SA budgets.

Table IV. Ring-artifact reconstruction accuracy before and after SEMO error mitigation under 15% projection sampling with 20% and 30% detector-row corruption. Energy reduction is defined as $\boldsymbol{\Delta E} = \boldsymbol{E_{SA}} - \boldsymbol{E_{SEMO}}$ for each paired run; positive values indicate lower QUBO energy after SEMO processing. Results for the 30%-corruption setting are reported as mean ± sample standard deviation across four random seeds (42, 123, 124, and 125).

| Corruption | Reads | Sweeps | Seed | SA wrong pixels | SEMO wrong pixels | Energy reduction, $\Delta E$ |
|---|---|---|---|---|---|---|
| 20% | 20 | 50 | 123 | 120 | 2 | 120.5 |
| 20% | 50 | 50 | 124 | 38 | 0 | 53.4 |
| 20% | 100 | 50 | 124 | 38 | 0 | 53.4 |
| 20% | 100 | 100 | 42 | 57 | 0 | 72.5 |
| 30% | 20 | 50 | 42,123,124,125 | 199.00 ± 9.56 | 173.50 ± 23.90 | 27.85 ± 15.13 |
| 30% | 50 | 50 | | 182.25 ± 21.08 | 170.00 ± 25.94 | 20.84 ± 10.01 |

Under 20% corruption, three of the four selected SA–SEMO pairs achieved exact reconstruction, while the remaining pair improved from 120 to 2 incorrectly reconstructed pixels. However, the 50-read and 100-read runs with 50 sweeps and seed 124 yielded identical initial SA bitstrings. Their SEMO outputs differed at the bit level but decoded to the same exact reconstruction; thus, the three exact-recovery pairs corresponded to two distinct initial SA configurations. Under 30% corruption, SEMO reduced the number of incorrectly reconstructed pixels in seven of the eight evaluated pairs, although none achieved exact reconstruction. In the run with 50 reads, 50 sweeps, and seed 42, the QUBO energy decreased by approximately 6.18, while the number of incorrectly reconstructed pixels increased from 186 to 187. Figure 3 summarizes the corresponding changes in QUBO energy and MAE across the main ROI-QCSTR and ring-artifact experiments.

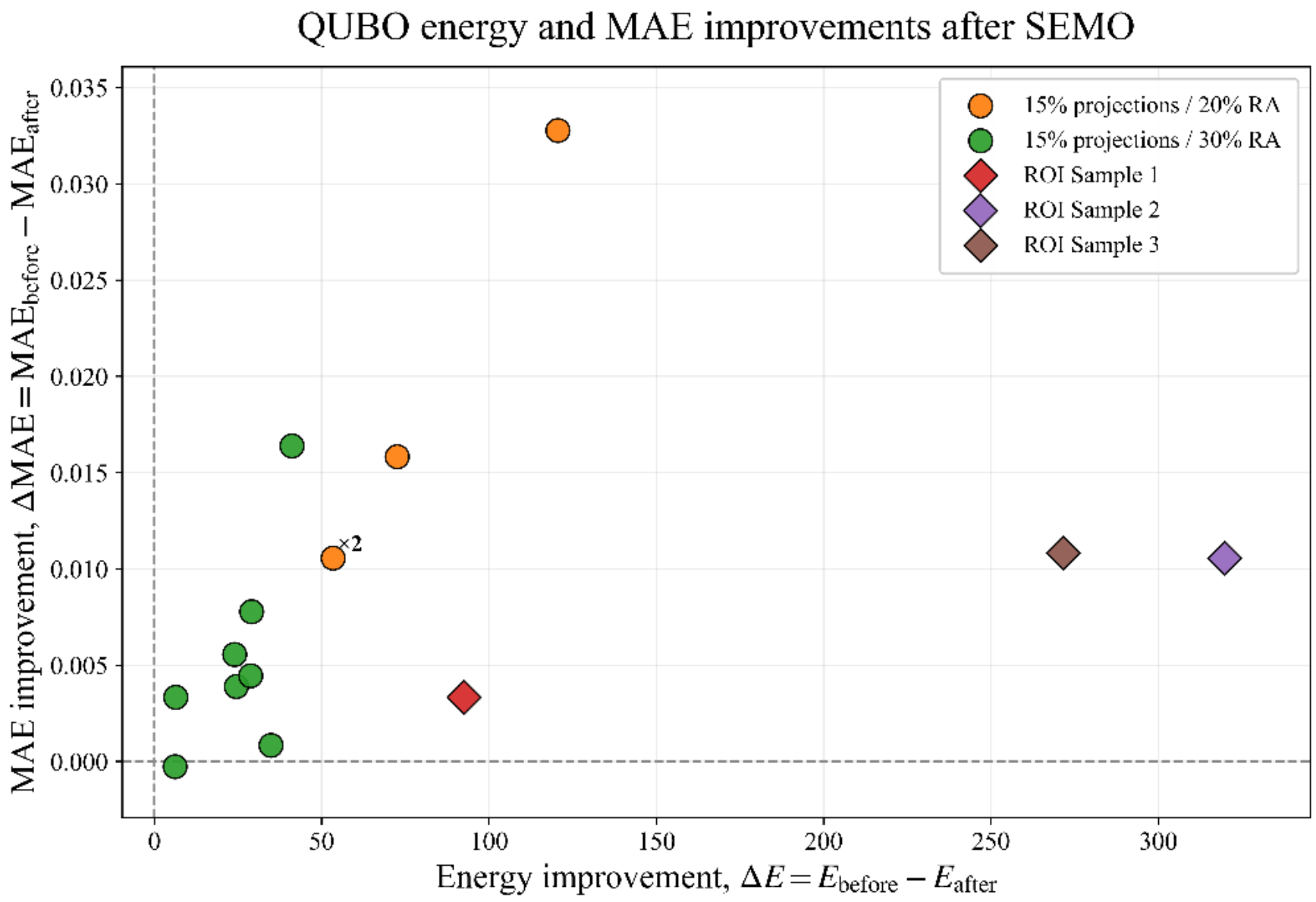

Figure 3. QUBO energy and MAE improvements after SEMO processing for the main ROI-QCSTR and ring-artifact experiments. The horizontal axis represents the reduction in QUBO energy, $\Delta E = E_{\mathbf{before}} - E_{\mathbf{after}}$, and the vertical axis represents the reduction in mean absolute error, $\Delta MAE = MAE_{\mathbf{before}} - MAE_{\mathbf{after}}$. Positive values on both axes indicate simultaneous improvement in the QUBO objective and image-domain reconstruction accuracy. The symbol $\times \mathbf{2}$ indicates two evaluated runs with identical $(\Delta E, \Delta MAE)$ coordinates.

### 3.4 SEMO error mitigation of hybrid-solver outputs

To examine whether SEMO error mitigation is applicable to solutions generated by different optimization methods, SEMO was also applied to candidate solutions obtained using the hybrid solver. The resulting reconstructions were evaluated using the same image-domain metrics as in the SA-based experiments.

In the 54-view ROI-QCSTR experiments, SEMO reduced the number of incorrectly reconstructed pixels in Sample 1 from 78 to 1. For Sample 2, the hybrid solver had already achieved exact reconstruction, and the number of incorrectly reconstructed pixels remained zero after SEMO processing. In Sample 3, both the original hybrid-solver solution and the SEMO-mitigated solution contained 15 incorrect pixels.

In the ring-artifact experiments, the hybrid solver had already achieved exact reconstruction under 15% projection sampling and 20% detector-row corruption, and the number of incorrectly reconstructed pixels remained zero after SEMO processing. Under 15% projection sampling with 30% detector-row corruption, SEMO reduced the number of incorrect pixels from 298 to 225.

## 4. Discussion

The present results provide an initial validation of the SA–SEMO workflow as a software-based strategy for improving quantum CT image reconstruction, motivated by the need to extend the computable ROI beyond the currently evaluated problem sizes. SEMO improved selected CT-QUBO solutions without modifying the initial optimizer or the reconstruction objective. Under the same low-budget SA setting, SEMO reduced reconstruction errors in all three 54-view ROI-QCSTR instances, achieving exact reconstruction in one case. It also achieved exact reconstruction from selected imperfect SA solutions in the sparse-view, corrupted-sinogram experiment and improved selected outputs of the hybrid solver. These findings establish the applicability of SEMO as a post-sampling refinement step across the two evaluated CT reconstruction settings and two types of initial optimizer. The magnitude of improvement, however, depended on the problem instance and the initial solution.

Notably, low-budget SA followed by SEMO attained the same number of incorrectly reconstructed pixels as all four evaluated higher-budget SA runs in ROI Samples 1 and 2. In Sample 3, it matched two of the four higher-budget SA results, although the best higher-budget run achieved fewer reconstruction errors. The SA–SEMO results also matched or improved upon the selected hybrid-solver outputs before SEMO processing in all three ROI instances. These observations suggest that post-sampling refinement may provide an alternative to increasing the initial SA sampling budget or using a hybrid solver to attain a given reconstruction accuracy in some CT-QUBO instances. Whether this approach offers an advantage in total execution time or computational cost remains to be established.

The results also indicate that improvements in the QUBO objective and image-domain reconstruction accuracy should not be treated as equivalent. Most evaluated cases showed reductions in both QUBO energy and MAE, but this relationship was not universal. For example, in one run under 30% detector-row corruption, SEMO reduced the QUBO energy by approximately 6.18 while increasing the number of incorrectly reconstructed pixels from 186 to 187. This behavior is consistent with the fact that the QUBO objective is defined through sinogram consistency and, for ROI-QCSTR, spatial regularization rather than direct comparison with the unknown ground-truth image. Multiple discrete reconstructions may therefore have similar objective values while differing in a small number of pixels. This distinction may become more pronounced when projection information is incomplete or corrupted, because the inverse problem is less constrained and different image configurations may be compatible with the available projection data.

ROI Sample 3 provides a concrete example of the distinction between QUBO energy and reconstruction accuracy. Although SEMO substantially improved the initial SA reconstruction, its remaining non-ground-truth configuration had lower energy than the exact target under the fixed QCSTR objective. The energy differences were $-21.067$ for the sinogram-consistency term and $-3.000$ for the spatial-regularization term. Correcting the residual pixel errors individually or together increased the total QUBO energy. These observations do not establish whether SEMO reached the global QUBO optimum or identify the source of the residual mismatch. Rather, they show that improving the specified CT-QUBO objective and recovering the exact target are distinct requirements.

A separate observation concerns the hybrid-solver result for ROI Sample 3, for which both the initial hybrid-solver solution and its SEMO-mitigated counterpart contained 15 incorrectly reconstructed pixels. Unlike the controlled SA experiments, the hybrid BQM solver does not expose the same read-, sweep-, and seed-level controls over its internal optimization process. Therefore, the present hybrid-solver experiment does not establish whether additional solver effort would have produced a different candidate solution. Moreover, the hybrid-solver configuration with 15 incorrect pixels has not been subjected to the same component-wise energy comparison performed for the SEMO solution obtained from low-budget SA. The hybrid result should therefore be interpreted separately and should not be assumed to share the same energy-landscape explanation without additional analysis.

A practical limitation of the present study is the restricted number of hybrid-solver evaluations. Repeated hybrid-solver runs were substantially more limited than local SA experiments, preventing a statistically comparable characterization of run-to-run variability and solution robustness. The hybrid-solver results should therefore be interpreted as representative comparisons rather than as an exhaustive benchmark of hybrid-solver performance. Additional repeated evaluations would be required to determine the reproducibility of the observed hybrid-solver solutions and the frequency with which SEMO provides further improvement. A controlled comparison of total execution time, including both initial sampling and SEMO processing, remains outside the scope of this study.

Beyond the reconstruction-quality improvements demonstrated in this study, an important motivation for developing the SA–SEMO workflow is to extend the spatial scale of quantum CT image reconstruction beyond the problem sizes evaluated to date in CT-QUBO studies using quantum and hybrid quantum-classical optimization. Under the four-level intensity representation used here, each pixel requires three binary variables; thus, the present 60 × 60 ROI contains 10,800 variables. Enlarging the ROI to 150 × 150 or 200 × 200 would require 67,500 or 120,000 variables, respectively. A problem size of approximately 100,000 binary variables would therefore accommodate an ROI of about 182 × 182 pixels, representing a threefold increase in linear dimension and more than nine times the pixel area of the current benchmark. To illustrate the spatial implications, consider a hypothetical 200-mm field of view represented by a 1000 × 1000 image, corresponding to a nominal in-plane sampling interval of 0.2 mm. Such a sampling interval is consistent with the reconstructed voxel size reported for a dedicated breast CT system [22]. Under this assumption, a 1-mm structure would span approximately five pixels along one dimension, although this sampling relationship alone does not establish its detectability, which also depends on system resolution, contrast, and noise. The current 60 × 60 ROI would cover only 12 × 12 mm, whereas 150 × 150 and 200 × 200 ROIs would cover 30 × 30 mm and 40 × 40 mm, respectively. Reconstructing the entire image using the same encoding would require three million binary variables, motivating progressive ROI refinement rather than direct whole-image optimization. Although the present experiments do not establish scalability to these larger problem sizes, they provide an initial solution-quality benchmark for subsequent development. Importantly, increasing the variable count also increases the potential number of pairwise QUBO interactions, making memory usage and objective evaluation additional scalability constraints. Future work will therefore investigate parallel SA–SEMO implementations using C/CUDA, together with memory-efficient QUBO representations and numerical-precision analysis, toward reconstruction problems involving 100,000 or more binary variables.

Taken together, these findings motivate further evaluation of SEMO as a complement to initial optimization and, in selected instances, as a possible alternative to increasing the initial sampling budget. The demonstrated improvements in SA-generated CT-QUBO solutions provide an initial basis for developing a software-based reconstruction framework aimed at extending the computable ROI beyond the current benchmark. Establishing its practical value and scalability will require controlled comparisons of reconstruction accuracy, total execution time, memory requirements, numerical precision, and computational cost at progressively larger problem sizes.

## 5. Conclusion

This study investigated SA combined with SEMO as an initial software-based strategy toward larger-scale quantum CT image reconstruction. Without modifying the underlying CT-QUBO objective, SEMO substantially improved low-budget SA reconstructions across three 10,800-variable ROI instances and selected sparse-view, corrupted-sinogram problems. In the ROI experiments, SA–SEMO achieved reconstruction accuracy comparable to or better than the selected unmitigated hybrid-solver outputs. Additional improvements were also observed when SEMO was applied directly to selected hybrid-solver solutions. However, the analysis of a non-exact ROI reconstruction showed that the fixed QUBO objective could assign lower energy to a reconstructed image than to the ground truth, highlighting the need to assess optimization energy and image-domain accuracy separately.

These findings provide an initial solution-quality benchmark for future development of a scalable software-based reconstruction pipeline, rather than demonstrating a computational speedup or larger-variable capability. Future work will investigate parallel SA–SEMO implementations using C/CUDA and memory-efficient QUBO representations, with the objective of extending reconstruction to problems involving 100,000 or more binary variables. Systematic evaluation of reconstruction accuracy, numerical precision, memory requirements, and end-to-end execution time will be essential for determining whether this approach can support substantially larger ROIs within an iterative quantum CT image reconstruction framework.

**Acknowledgments**

The D-Wave Leap hybrid-solver computational time used in this study was provided through the project "Creation of the Quantum Information Science R&D Ecosystem (Based on Human Resources)," funded by the Korean government through the National Research Foundation of Korea (NRF) (RS-2023-NR057243). H. Lee was supported by an NRF grant funded by the Ministry of Science and ICT (MSIT), Republic of Korea (RS-2024-00352408). The authors would like to acknowledge discussions with Peter Tyson, Chris Vale and Anthony Chesman of CSIRO. The Python-compatible SEMO binary software module used in this article is available at http://data.csiro.au.

**Conflict of Interest Declaration**

Two of the authors (YSY and ABM) are co-inventors of, and CSIRO owns, the pending patent [14] on which the SEMO method is based.